\documentclass[12pt,english,floatfix,superscriptaddress,aps,prd,preprint,showkeys,nofootinbib]{revtex4}
\usepackage{amsmath}
\usepackage{amssymb}
\usepackage{amsbsy}
\usepackage{amsfonts}
\usepackage{amsopn}
\usepackage{amstext}
\usepackage{graphicx}
\usepackage[english]{babel}
\usepackage{color}
\usepackage{slashed}
\usepackage{esint}
\usepackage[dvips]{epsfig}
\usepackage[dvips]{graphicx}
\usepackage{float}
\usepackage{units}
\usepackage{textcomp}
\usepackage{wasysym}
\usepackage{hyperref}
\usepackage{caption}
\usepackage{subcaption}
\begin{document}

\title{Most likely configurations for fermion localization in a Braneworld-$f(Q,B_Q)$ }

\author{A. R. P. Moreira}
\email{allan.moreira@fisica.ufc.br}
\affiliation{Reserach Center for Quantum Physics, Huzhou University, Huzhou, 313000, P. R. China.}
\affiliation{Secretaria da Educaç\~{a}o do Cear\'{a} (SEDUC), Coordenadoria Regional de Desenvolvimento da Educaç\~{a}o (CREDE 9),  Horizonte, Cear\'{a}, 62880-384, Brazil}
\author{Shi-Hai Dong}
\email{dongsh2@yahoo.com}
\affiliation{Reserach Center for Quantum Physics, Huzhou University, Huzhou, 313000, P. R. China.}
\affiliation{Centro de Investigaci\'{o}n en Computaci\'{o}n, Instituto Polit\'{e}cnico Nacional, UPALM, CDMX 07700, Mexico}
\author{Manuel E. Rodrigues}
\email{esialg@gmail.com}
\affiliation{Faculdade de Ci\^{e}ncias Exatas e Tecnologia, Universidade Federal do Par\'{a}, Campus Universit\'{a}rio de Abaetetuba, 68440-000, Abaetetuba, Par\'{a}, Brazil}
\affiliation{Faculdade de F\'{\i}sica, Programa de P\'{o}s-Graduaç\~{a}o em F\'{\i}sica,
Universidade Federal do Par\'{a}, 66075-110, Bel\'{e}m, Par\'{a}, Brazil}

\begin{abstract}

This study delves deeply into braneworld scenarios within modified gravity models, investigating their impact on particle localization and the structure of branes. Through a comprehensive blend of numerical analyses and theoretical inquiries, we unravel a nuanced correlation between deviations from standard General Relativity (GR) and the emergence of split branes. By employing probabilistic measurements, we pinpoint stable configurations that align with brane division intervals, thus challenging prevailing assumptions regarding the gravitational framework of our universe. Furthermore, our investigation extends to the localization of fermions within the brane, exposing intricate dynamics shaped by scalar field characteristics and modifications to gravitational models. By harnessing quantum information measurements, notably Shannon entropy, we discern heightened probabilities of fermion localization within the brane as gravitational models diverge from standard paradigms. This underscores the limitations of General Relativity in comprehensively describing the complexities inherent in our universe. Lastly, our exploration of massive fermions unveils their potential to breach the confines of the brane, hinting at promising avenues for future experimental endeavors aimed at probing the nature of extra dimensions and gravitational interactions. This suggests exciting prospects for advancing our understanding of fundamental physics beyond conventional boundaries.
\end{abstract}
\maketitle

\section{Introduction}

Currently, there is a growing fascination with exploring alternative gravity models, a curiosity that echoes back to the early days of GR. Despite the groundbreaking insights provided by GR, persistent gaps in our understanding have remained since its inception. Modern physics faces numerous enigmas, including the perplexing accelerated expansion of the universe \cite{Perlmutter1999, Riess1999, Gonzalez-Gaitan2011, Ganeshalingam2011,Nojiri:2017ncd}, the elusive nature of dark matter, which defies explanation within the Standard Model \cite{Boehm:2000gq,SupernovaSearchTeam:1998fmf}, the hierarchy problem \cite{rs,rs2}, and the mechanisms governing baryonic symmetry in the cosmos. These unresolved puzzles serve as catalysts for exploring scenarios that transcend the boundaries of GR.

The braneworld concept has led to significant theoretical advancements in addressing several of these persistent mysteries of GR \cite{Gremm1999,CastilloFelisola2004,Navarro2004,BarbosaCendejas2005,Bazeia2007,Liu2011wi,fR1,fR2,tensorperturbations,ftnoncanonicalscalar,ftborninfeld,ftmimetic,Belchior:2023xgn, Moreira:2023uys,Moreira:2023pes,Belchior:2023gmr}. Furthermore, this theory predicts the existence of a $750\ GeV$ particle, which has appeared a few times at the Large Hadron Collider but was identified as a failed measurement \cite{Arun:2017zap,Arun:2016ela,Arun:2015ubr}.

Within the realm of alternative gravitational theories, attention has gravitated toward diverse avenues beyond GR. For instance, Einstein-Cartan geometry \cite{Hehl:1976kj}, and metric-based models like $f(R)$ theories \cite{DeFelice:2010aj}, have piqued interest as potential deviations from the standard framework. Another intriguing approach is found in the teleparallel equivalent of general relativity (TEGR), which postulates gravity as a consequence of spacetime torsion rather than curvature \cite{Aldrovandi}. In this model, the \textit{vielbein} field serves as a dynamic variable, operating under the assumption of the absence of the Riemann curvature tensor.

In more recent developments, the Symmetric Teleparallel Equivalent of General Relativity (STEGR) has emerged as a compelling alternative \cite{Nester:1998mp}. STEGR introduces the non-metricity tensor into the dynamics of gravitational degrees of freedom, setting it apart from TEGR. Variants such as the $f(Q)$ gravity model have been proposed, offering increased degrees of freedom compared to GR, contingent upon coefficients within the Lagrangian \cite{Hohmann:2018wxu,Soudi:2018dhv,BeltranJimenez:2017tkd,BeltranJimenez:2019esp,Bajardi:2020fxh, Capozziello:2022tvvi,Capozziello:2022wgl,BeltranJimenez:2019tme,Bhar:2023zwi,Atayde:2023aoj,Koussour:2023rly,Bajardi:2023vcc,Lin:2021uqa,Mustafa:2021ykn}.

Moreover, a gravitational model that has garnered considerable attention is gravity $f(Q,B_Q)$. This model has yielded significant results in addressing some of the unresolved puzzles within the framework of GR. In reference \cite{39,39.1,39.2}, a study was undertaken on FLRW cosmology within the framework of the $f(Q,B_Q)$ theory, exploring various families of connections. Additionally, investigations into the behavior of cosmological models of dark energy, as described by the same theory, were conducted in references \cite{40.1} using perfect fluid, and in reference \cite{2.1} using quintessence. However, there remains much to be explored about this model. Hence, we are motivated to investigate a scenario involving extra dimensions. This work represents the first endeavor in the literature to examine the braneworld scenario within this gravitational model.


The structure of this paper unfolds as follows: Section \ref{s1} introduces the fundamental concepts crucial for constructing the braneworld within the framework of gravity $f(Q,B_Q)$. Moreover, it identifies the most probable configurations and assesses the stability of the model. In Section \ref{s2}, the conditions governing the localization of fermions on the brane are scrutinized. Additionally, configurations with the highest likelihood of detecting massless fermions on the brane are investigated. Finally, our concluding remarks and future outlook are presented in Section \ref{s3}.

\section{Braneworld - $f(Q,B_Q)$}
\label{s1}

In this section, we will commence by elucidating the fundamental principles of symmetric teleparallel gravity and outlining the equation of motion relevant to $f(Q,B_Q)$ gravity. When navigating through theories entrenched in Riemannian geometry, it becomes imperative to adhere to the metricity condition $\nabla_M g_{NP}=0$. Here, $g_{NP}$ represents the metric, while $\nabla_M$ denotes the covariant derivative employing the Levi-Civita $\Gamma^P\ _{MN}$ as the affine connection. Notably, this condition aligns with the framework of GR (GR). However, the same cannot be said for theories grounded in non-Riemannian geometry. Within this domain, noteworthy among the modified gravity models is STEGR, distinguished by the presence of a nonvanishing nonmetricity tensor \cite{Nester:1998mp}
\begin{equation}
Q_{MNP}\equiv\nabla_M g_{NP}\neq 0.   
\end{equation}
To characterize this tensor, let us delineate its independent traces, namely 
\begin{eqnarray}
Q_M&=&g^{NP}Q_{MNP}=Q_{M}\ ^P\ _P,\nonumber\\  \widetilde{Q}_M&=&g^{NP}Q_{NMP}=Q^{P}\ _{MP}.   
\end{eqnarray}
Moreover, given the existence of the nonmetricity tensor, it necessitates the establishment of a broader connection 
\begin{equation}
\widetilde{\Gamma}^P\ _{MN}=\Gamma^P\ _{MN}+L^P\ _{MN}.   
\end{equation}
The entity termed $L^P\ _{MN}$ is commonly referred to as the distortion tensor. It is expressed in terms of the nonmetricity tensor \cite{Nester:1998mp}
\begin{equation}
L^P\ _{MN}=\frac{1}{2}g^{PQ}(Q_{PMN}-Q_{MPN}-Q_{NPM}).    
\end{equation}

In formulating a gravitational action tailored for STEGR, we introduce a comprehensive tensor that encapsulates not only the nonmetricity but also its autonomous traces and the distortion tensor. This tensor, termed the nonmetricity conjugate, plays a pivotal role within this framework \cite{Nester:1998mp}
\begin{equation}
P^P\ _{MN}=-\frac{1}{2}L^P\ _{MN}+\frac{1}{4}(Q^P-\widetilde{Q}^P)g_{MN}-\frac{1}{8}(\delta^P_M Q_N+\delta^P_N Q_M).    
\end{equation}
Furthermore, upon contraction with the nonmetricity tensor, this entity yields the nonmetricity scalar
\begin{equation}
Q=Q_{PMN}P^{PMN}.    
\end{equation}

Expressed as $R=Q+B_Q$, where
\begin{equation}
B_Q=\nabla_M(Q^M-\widetilde{Q}^M),   
\end{equation}
being the boundary term, the Ricci scalar underscores a notable facet of STEGR. This formulation indicates an equivalence between STEGR and GR, as the boundary term dissipates upon integration into the action. However, such parity does not extend to $f(Q)$ and $f(R)$ gravities due to the persistence of the boundary term. Nevertheless, an equivalence to gravity $f(R)$ can be achieved if gravity $f(R)$ is considered as $f(Q,B_Q)$.

This study is primarily focused on $f(Q,B_Q)$ gravity, representing yet another plausible extension of STEGR
\begin{equation}\label{a1}
S=\int d^5x \sqrt{-g}\Big[f(Q,B_Q)+2\kappa^2\mathcal{L}_m\Big],
\end{equation}
where $\mathcal{L}_m$ represents the matter lagrangian.

Variating the action (\ref{a1}) with respect to the metric yields the following amended Einstein equation
\begin{eqnarray}
f_Q G_{MN}-\frac{1}{2}g_{MN}\Big(f-f_Q-f_B B_Q\Big)+2P^P\ _{MN}\partial _P \Big(f_Q+f_B\Big)&&\nonumber\\ -g_{MN}\Box f_B+\nabla_M \nabla_N f_B&=&\kappa^2 \mathcal{T}_{MN},  
\end{eqnarray}
where
\begin{equation}
\mathcal{T}_{MN}=-2\frac{\delta \mathcal{L}_m}{\delta g^{MN}}+ g_{MN}\mathcal{L}_m,
\end{equation}
is the momentum-energy tensor. Furthermore, we denote $f$ as shorthand for $f(Q,B_Q)$, $f_Q$ as the partial derivative of $f$ with respect to $Q$, and $f_B$ as the partial derivative of $f$ with respect to $B_Q$, for the sake of simplicity.

Conversely, when adjusting the action (\ref{a1}) to enhance its connection, we aim to achieve
\begin{equation}
\nabla_M\nabla_N\Big[\sqrt{-g} P_K\ ^{MN}(f_Q+f_B)\Big]=0.   
\end{equation}

For an extra dimensions scenario, we will use the Randall-Sundrum-like metric
\begin{equation}\label{metric}
ds^2= e^{2A}\eta_{\mu\nu}dx^{\mu}dx^{\nu}+dy^2.    
\end{equation}
Here, $\eta_{\mu\nu}$ represents the Minkowski metric (the familiar 4D spacetime where we reside), $e^{2A}$ denotes the warp factor, and $y$ represents the extra dimension. The uppercase Latin indices $M, N$ range from 0 to 4, denoting the bulk coordinate indices (in the 5D spacetime). The Greek indices $\mu, \nu$, ranging from 0 to 3, correspond to the indices on the brane.

Furthermore, we will only consider a background scalar field capable of generating the brane, with a laragian described as
\begin{equation}\label{a3}
\mathcal{L}_m=-\Big[\frac{1}{2}\nabla_M\phi\nabla^M\phi+V(\phi)\Big],  
\end{equation}
where $\phi(y)$ is the scalar field which similar to the warp factor depends only on the extra dimension.

Consequently, for this metric, the scalar field and gravitational equations can be expressed as follows:
\begin{eqnarray}\label{erer}
\phi^{\prime\prime}+4A^{\prime}\phi^{\prime}&=&\frac{dV}{d\phi},\nonumber\\
24A'\Big[A'''(f_{BB}+f_{BQ})+A'A''(8f_{BB}+11f_{BQ}+3f_{QQ})\Big]&& \nonumber\\ \frac{1}{2}f+(4f_B-3f_Q)(4A'^2+A'')+3A'f'_{B}+f''_{B}&=&\kappa ^2\Big(\frac{1}{2}\phi'^2+V\Big),\nonumber\\
-\frac{1}{2}f-4\Big[A'f'_B+f_B(A''+4A'^2)-3A'^2f_Q\Big]&=&\kappa^2\Big(\frac{1}{2} \phi'^2- V\Big).
\end{eqnarray}
In the equations above, the prime ($'$) denotes the derivative with respect to the extra dimension.

We need to define the form of the function $f(Q,B_Q)$, as it will describe our gravitational model. To ensure generality, we will consider two models that generalize STEGR, namely, $f_1(Q,B_Q)=Q+kB_Q^n$ and $f_2(Q,B_Q)=Q+k_1B_Q+k_2B_Q^2$. Here, the parameters $n$ and $k$ determine the deviation from the usual STEGR. These models become particularly interesting because we can recover STEGR when $k=k_1=k_2=0$. Conversely, when $n=k=k_1=1$ and $k_2=0$, we arrive at GR, i.e., $f_{1,2}(Q,B_Q)=R$. Furthermore, since our focus is on studying the localization of fermions on thick branes, we need to define the ansatz for the warp factor that best characterizes a thick brane. Therefore, we choose
\begin{align}
\label{coreA}
A(y)=-p\ln{\cosh(\lambda y)},
\end{align}
where the parameters $\lambda$ and $p$ control the width of the brane. This ans\"{a}tz presents a more acceptable configuration of our universe \cite{Gremm1999,Moreira:2023pes}. 

With all these results in hand, we can finally embark on our study of the behavior of the braneworld scenario within these modified gravity models. To guide us, we can propose some basic questions and endeavor to address them comprehensively. Firstly, how does this gravitational change affect the structure of the brane? Does it modify the behavior of the background scalar field that generates the brane? Secondly, what is the most likely gravitational configuration to be found in nature, i.e., one that describes our world accurately? Thirdly, considering that the most likely way to experimentally prove the existence of extra dimensions is through theoretical predictions of particle locations on the brane, how do gravitational changes affect the locations of these particles, such as spin-$1/2$ fermions? Does the most probable configuration of the gravitational model align with the configuration offering the greatest chances of locating the particle on the brane? Fourthly, is there any possibility of the particle escaping the brane and then returning, as observed experimentally with a particle of $750\ GeV$ \cite{Arun:2017zap,Arun:2016ela,Arun:2015ubr}?

To begin, we analyze the energy densities of each gravitational model. Energy density is defined in the form:
\begin{align}
\label{coreA}
\rho(y)=-e^{2A}\mathcal{L}_m.
\end{align}
We plot the energy density behavior for the first gravitational model in Fig. \ref{fig1}. Note that for $n=1$, the energy density exhibits a division at the peak. A single peak bifurcates into two as we deviate from STEGR (by increasing the influence of the parameter $k$). This division occurs within the range $0.20<k<0.26$ (Fig. \ref{fig1}a). A similar phenomenon occurs for $n=2$, albeit the division occurs in an interval closer to STEGR, specifically $-0.006<k<-0.004$ (Fig. \ref{fig1}b). For $n=3$, this interval becomes even narrower, $0.00004<k<0.00010$ (Fig. \ref{fig1}c). We can infer that the higher the power in $B_Q$, the more rapidly the energy density peak of the model tends to split.

\begin{center}
\begin{figure}[ht!]
\begin{centering}
\subfloat[$n=1$]{\centering{}\includegraphics[scale=0.6]{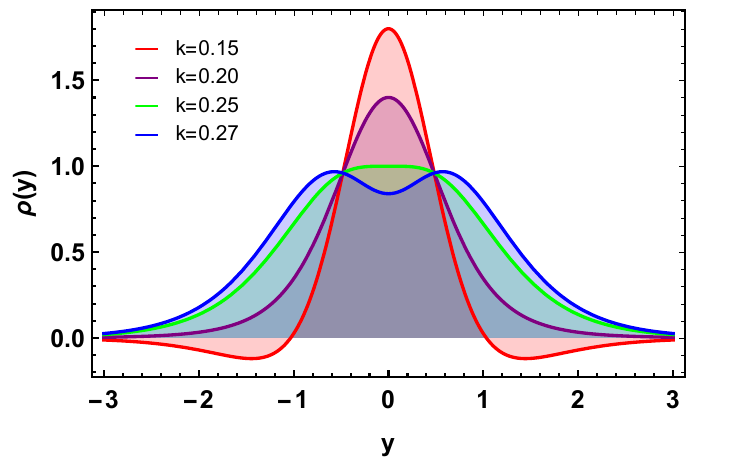}}
\subfloat[$n=2$]{\centering{}\includegraphics[scale=0.6]{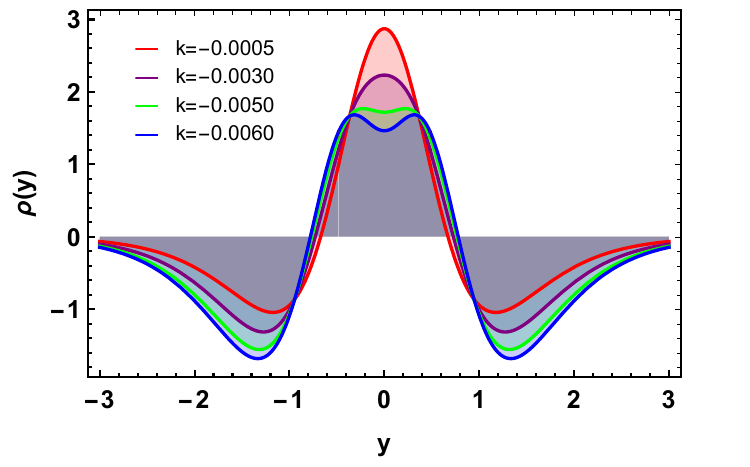}}
\par\end{centering}
\begin{centering}
\subfloat[$n=3$]{\centering{}\includegraphics[scale=0.6]{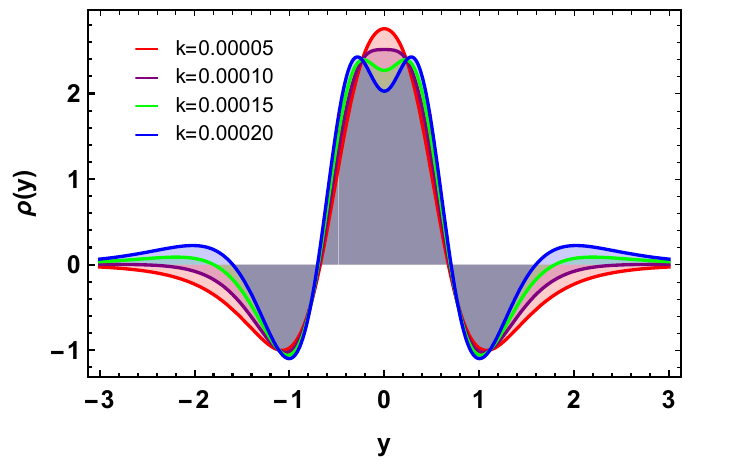}}
\par\end{centering}
\centering{}\caption{The energy density for $f_{1}(Q,B_Q)$ with $\kappa=\lambda=p=1$}\label{fig1}
\end{figure}
\par\end{center}

In Fig. \ref{fig2}, we illustrate the behavior of the energy density for the second gravitational model. As anticipated, varying the parameters $k_{1,2}$ leads to a division of the energy density peak. When setting the value of $k_2=-0.001$, we observe that the division occurs within the range $0.15<k_1<0.25$. Similarly, when setting $k_1=0.01$, the division takes place between $-0.005<k_2<-0.003$. Notably, the greater the deviation from STEGR, the more rapidly the energy density tends to split into two peaks.

\begin{center}
\begin{figure}[ht!]
\begin{centering}
\subfloat[$k_2=-0.001$]{\centering{}\includegraphics[scale=0.6]{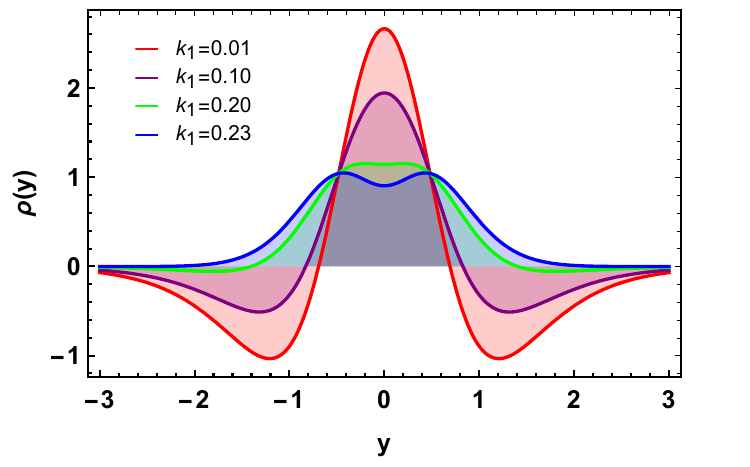}}
\subfloat[$k_1=0.01$]{\centering{}\includegraphics[scale=0.6]{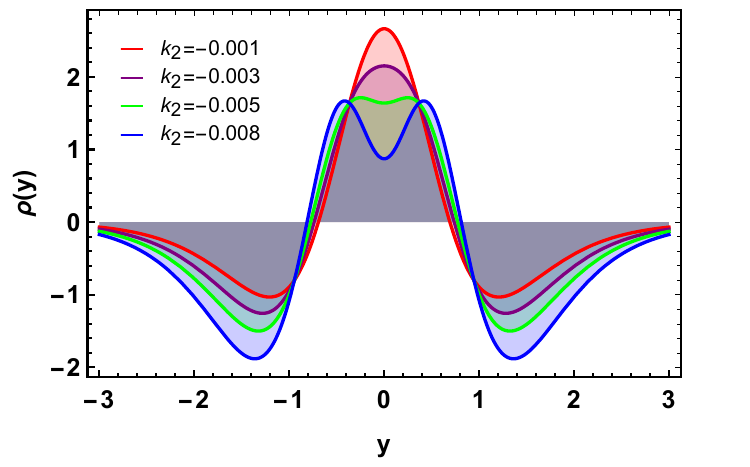}}
\par\end{centering}
\centering{}\caption{The energy density for $f_{2}(Q,B_Q)$ with $\kappa=\lambda=p=1$}\label{fig2}
\end{figure}
\par\end{center}

The intriguing aspect is that this division in the energy density signifies a split in the brane, indicating the emergence of a parallel world. In other words, as our models deviate further from the standard STEGR, the split in the brane becomes more pronounced. In GR, this phenomenon occurs when two background fields capable of generating the brane are added. However, in our models, we were able to observe a very pronounced split of the brane with just a single scalar field in the background.

To delve deeper, we can identify which configurations are most probable in our models. To accomplish this, we employ a powerful mathematical tool known as Differential Configurational Entropy (DCE) \cite{GS}. DCE, a probabilistic measure grounded in information theory, facilitates the determination of the most stable configurations of the model. These stable configurations offer insights into the most likely configurations of the model \cite{Correa2015a,Correa2015c,Correa2015b,Correa2016b,Correa2016a,Moreira:2021cta,Moreira:2022zmx}.

The DCE is constructed upon the energy density of the brane, which is mathematically represented as follows:
\begin{align}\label{dce_g}
S_C[f]=-\int \bar{f}(\omega)\ln[\bar{f}(\omega)]d\omega.
\end{align}
In this equation, $\bar{f}(\omega)$ denotes a normalized function of the frequency $\omega$. This function is defined as follows:
\begin{align}
f(\omega)=\frac{\mid\mathcal{F}[\omega]\mid^2}{\int\mid\mathcal{F}[\omega]\mid^2d\omega},
\end{align}
Here, $f(\omega)$ denotes the modal fraction, quantifying the relative weight of each mode $\omega$ ($\leq 1$). The term $\mathcal{F}[\omega]$ represents the Fourier transform of the energy density and is defined as follows:
\begin{align}
\mathcal{F}[\omega]=\frac{1}{\sqrt{2\pi}}\int e^{i\omega y}\rho(y) dy.
\end{align}
It is crucial to note that $\bar{f}(\omega)=f(\omega)/f_{\text{max}}(\omega)$, where $f_{\text{max}}(\omega)$ represents the maximum value of the modal fraction. Additionally, it is pertinent to highlight that the formulation (\ref{dce_g}) holds validity solely for continuous functions within an open interval.

For the first gravitational model, we numerically plot the shape of the modal fraction and the DCE in Fig. \ref{fig3}. In Fig. \ref{fig3}(a), it is evident that the modal fraction undergoes several modifications in its structure as we vary the value of $k$. This variation is linked to the deviation from the usual STEGR. The most intriguing aspect is demonstrated in the DCE plot. The steepest maximum corresponds to a trough. This valley signifies the most stable configuration of the model. The most likely configuration falls within the range of $0.20<k<0.26$. It's noteworthy that this interval aligns with the one identified in the energy density as the landmark for the split in the brane. The same pattern emerges for $n=2$ in the range $-0.006<k<-0.004$ (Fig. \ref{fig3}(b)), and for $n=3$ in the range $0.00004<k<0.00010$ (Fig. \ref{fig3}(c)).

\begin{center}
\begin{figure}[ht!]
\begin{centering}
\subfloat[$n=1$]{\centering{}\includegraphics[scale=0.6]{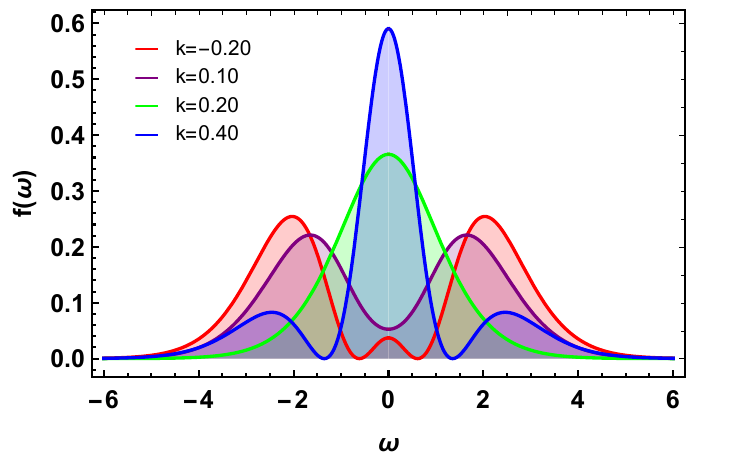}
\includegraphics[scale=0.56]{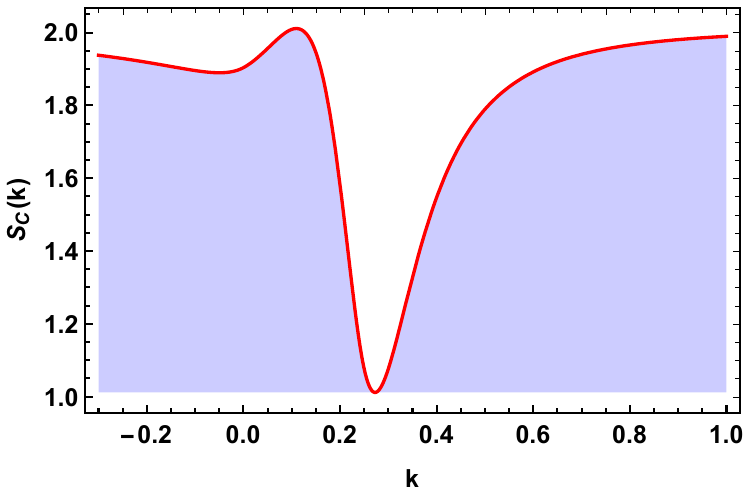}}
\par\end{centering}
\begin{centering}
\subfloat[$n=2$]{\centering{}\includegraphics[scale=0.6]{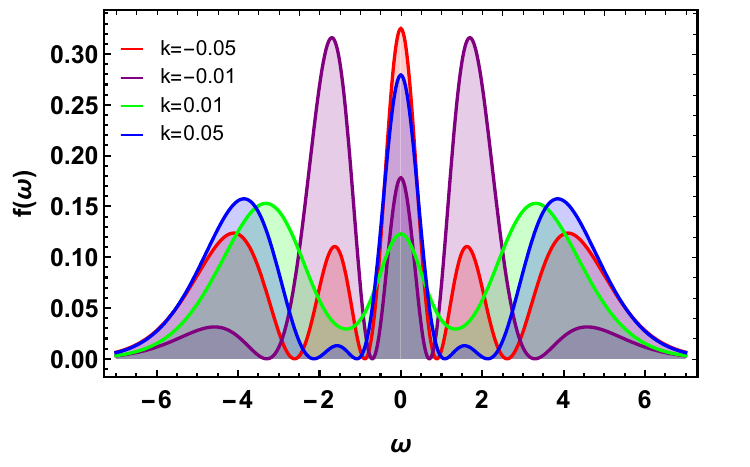}\includegraphics[scale=0.56]{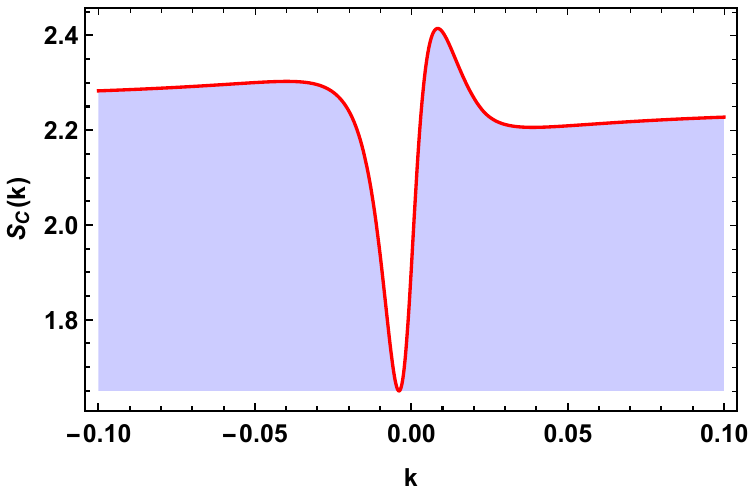}}
\par\end{centering}
\begin{centering}
\subfloat[$n=3$]{\centering{}\includegraphics[scale=0.6]{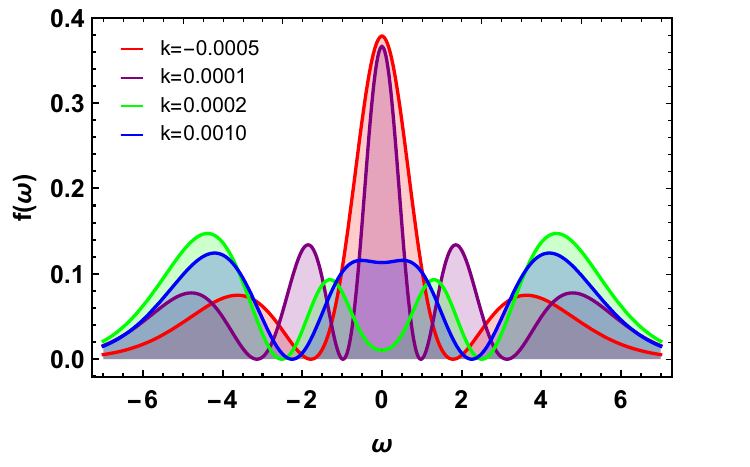}\includegraphics[scale=0.56]{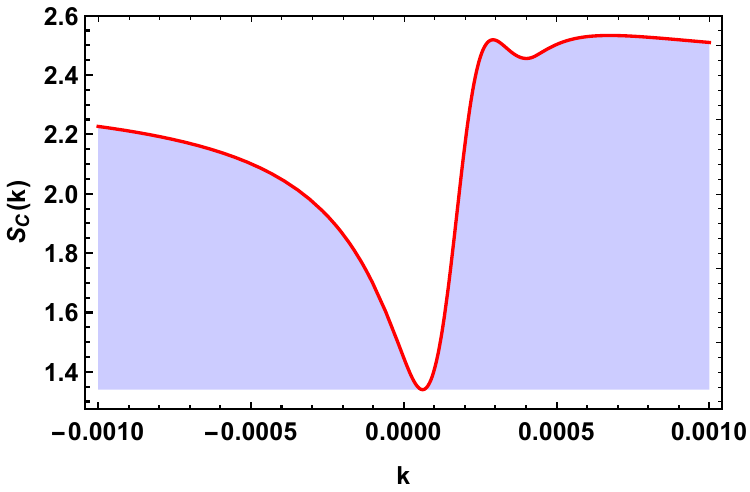}}
\par\end{centering}
\centering{}\caption{Modal fraction and DCE for $f_{1}(Q,B_Q)$ with $\kappa=\lambda=p=1$}\label{fig3}
\end{figure}
\par\end{center}

For the second gravitational model, the numerical plots depicted in Fig. \ref{fig4} illustrate the behavior of the modal fraction and the DCE. When we set the value of $k_2=-0.001$, we observe that the valley in the DCE plot appears in the range of $0.15<k_1<0.25$ (Fig. \ref{fig4}a). Similarly, when we set $k_1=0.01$, the valley emerges in the range of $-0.005<k_2<-0.003$ (Fig. \ref{fig4}b). Once again, these intervals coincide with those marking the split in the brane. We can thus conclude that the most stable configurations of our models are those that coincide with the appearance of the split in the brane. This is evident in both $f_1$ and $f_2$. Moreover, these configurations are the most likely ones to describe the universe in which we live. This result is significant as it assures us that the usual STEGR is not the most likely configuration to describe our world. In other words, we do not inhabit the usual STEGR. Or rather, we do not live in the standard GR!

\begin{center}
\begin{figure}[ht!]
\begin{centering}
\subfloat[$k_2=-0.001$]{\centering{}\includegraphics[scale=0.6]{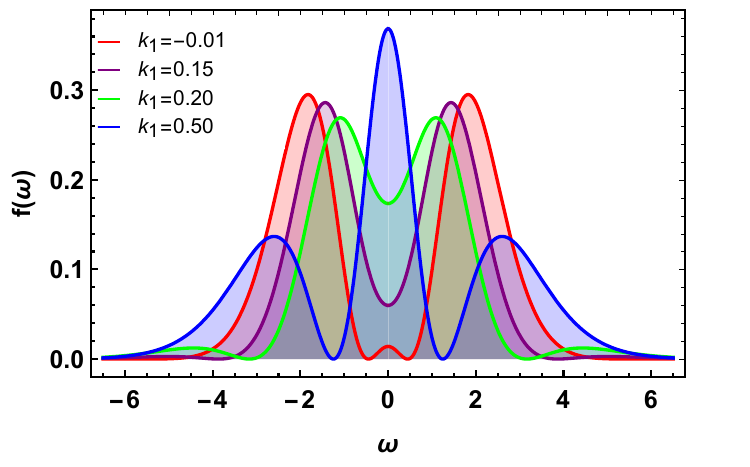}\includegraphics[scale=0.56]{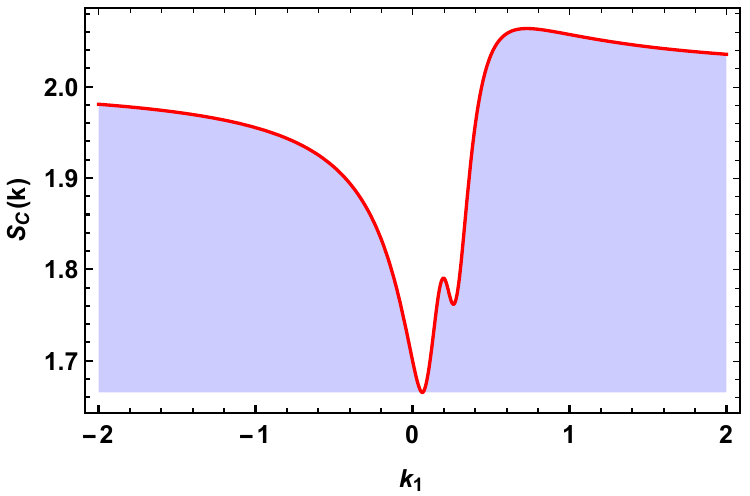}}
\par\end{centering}
\begin{centering}
\subfloat[$k_1=0.01$]{\centering{}\includegraphics[scale=0.6]{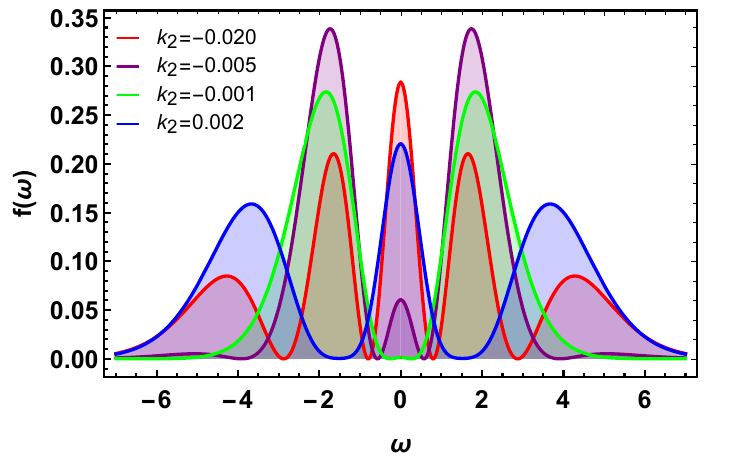}\includegraphics[scale=0.56]{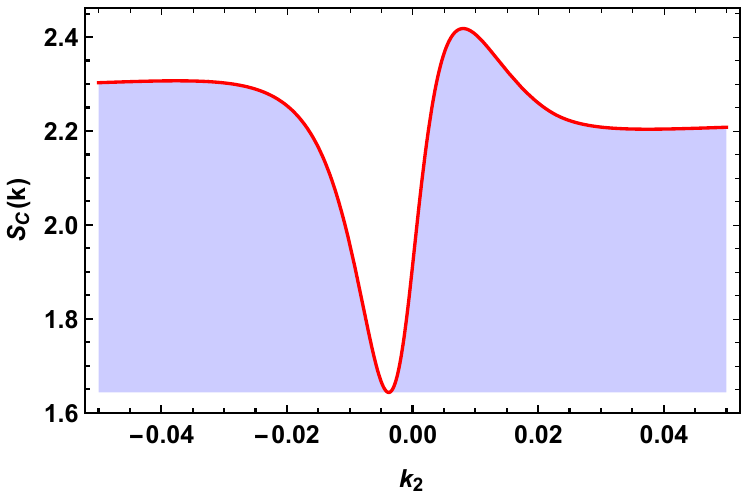}}
\par\end{centering}
\centering{}\caption{Modal fraction and DCE for $f_{2}(Q,B_Q)$ with $\kappa=\lambda=p=1$}\label{fig4}
\end{figure}
\par\end{center}

\section{Fermion localization mechanism}
\label{s2}

We go even further in our theoretical analyses. In this section, we will study the possibility of locating the particle in the brane. For this, let's consider spin-$1/2$ fermions.

To facilitate fermion localization, we need to introduce a coupling to the spinor field. The simplest and most efficient coupling is the well-known Yukawa coupling, where the spinor is minimally coupled with the scalar field $\phi$. Therefore, the scalar field solution must satisfy the following basic requirements to enable fermion localization in the brane:
\begin{enumerate}
    \item The field $\phi$ must be asymmetric at the origin of the brane. This represents a phase transition constrained by the structure of the brane.
     \item Asymptotically the field $\phi$ must tend to a constant, i.e., $\phi(y\rightarrow\pm\infty)\rightarrow\pm\phi_c$. Thus, $\partial V(\phi\rightarrow\pm\phi_c)/\partial\phi=0$. This guarantees the physical sense of the model.
\end{enumerate}

To obtain the value of the scalar field we use Eq.(\ref{erer}), which leads to the first-order differential equation:
\begin{eqnarray}\label{erer2}
\phi^{\prime}=\frac{1}{\kappa^2}\Big\{24(8f_{BB}+11f_{BQ}+3f_{QQ})A'^2A''+f_{BB}''-3f_QA''+2A'[f_B'+12(f_{BB}+f_{BQ})A''']\Big\}.\nonumber\\
\end{eqnarray}
However, the solution to Eq.(\ref{erer2}) is not so simple. This will depend on our choice for the form of $f(Q,B_Q)$. Therefore, our analysis is done numerically.

In Fig. \ref{fig5}, we plot the scalar field solution for $n=2$ and $3$ of $f_1$. These solutions are referred to as kink-type solutions. They are not precisely topological kinks, but they exhibit similar behavior. Noticeably, for a gravitational model close to STEGR, the scalar field solution presents only one domain wall. These outcomes are anticipated. However, something unusual occurs when we modify our gravitational model. For $n=2$, as we increase the value of $k$, the emergence of new domain walls becomes evident. This causes our field solution to undergo a phase transition, transitioning from a kink-type solution to a double-kink-type solution. This transition is linked to the brane split that occurs in the interval $-0.006<k<-0.004$. Similarly, the same phenomenon occurs for $n=3$ in the range $0.00004<k<0.00010$.
\begin{center}
\begin{figure}[ht!]
\begin{centering}
\subfloat[$n=2$]{\centering{}\includegraphics[scale=0.55]{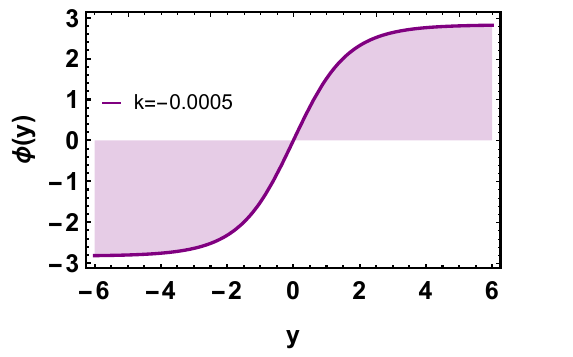}
\includegraphics[scale=0.55]{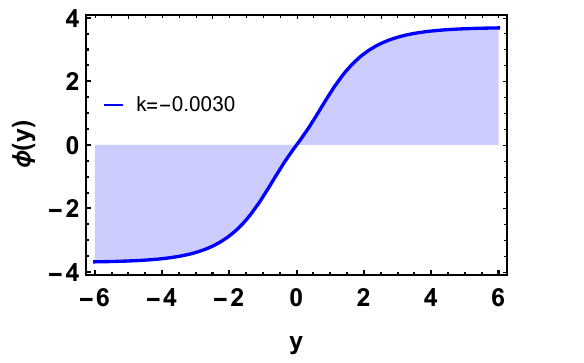}\includegraphics[scale=0.54]{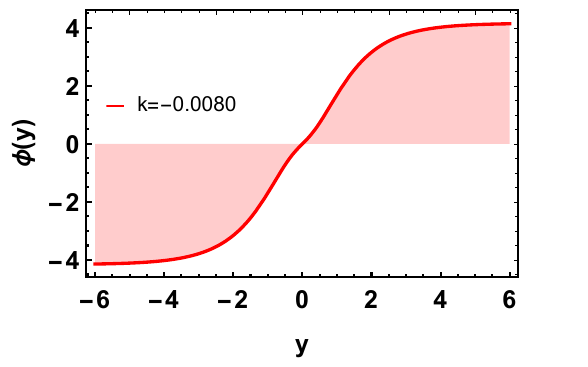}}
\par\end{centering}
\begin{centering}
\subfloat[$n=3$]{\centering{}\includegraphics[scale=0.55]{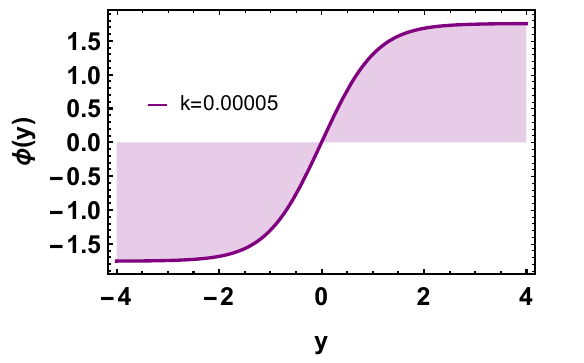}
\includegraphics[scale=0.52]{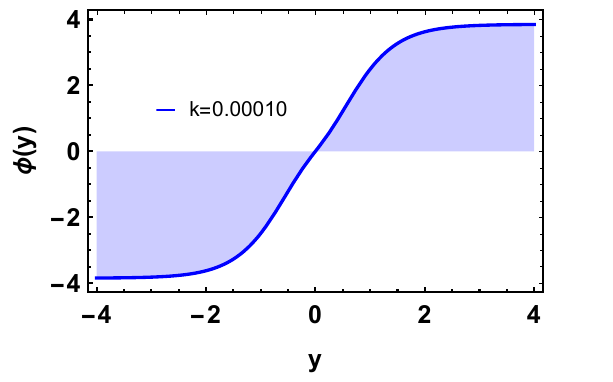} \includegraphics[scale=0.52]{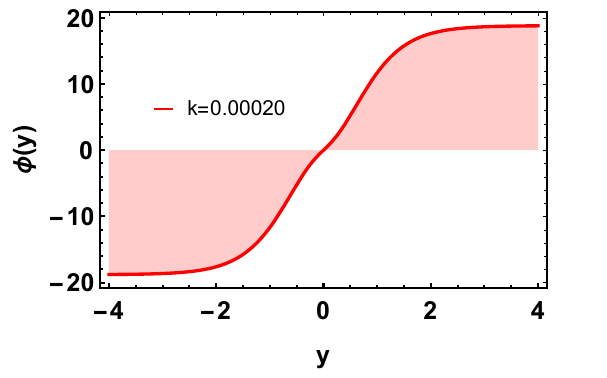}}
\par\end{centering}
\centering{}\caption{Double-kink type solution for $f_{1}(Q,B_Q)$ with $\kappa=\lambda=p=1$}\label{fig5}
\end{figure}
\par\end{center}

For $f_2$, we present the scalar field solution in Figure \ref{fig6}. It is evident that as we deviate our gravitational model from the usual STEGR, new structures emerge in the brane. New domain walls appear, transforming our kink-type solution into a double-kink-type solution as we vary from $0.15<k_1<0.25$ to $k_2=-0.001$, and $-0.005<k_2<-0.003$ to $ k_1=0.01$. Notably, this interval coincides with that of the brane split, indicating that the background field senses the split in the brane.
\begin{center}
\begin{figure}[ht!]
\begin{centering}
{\centering{}\includegraphics[scale=0.54]{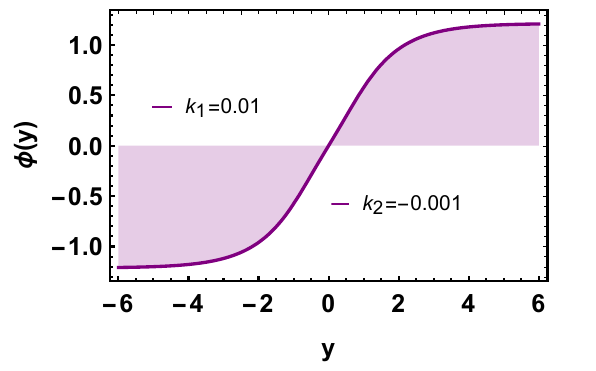}
\includegraphics[scale=0.55]{fig5b.pdf} \includegraphics[scale=0.49]{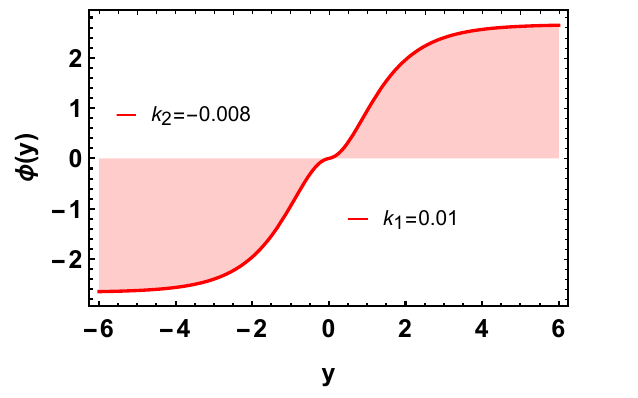}}
\par\end{centering}
\centering{}\caption{Double-kink type solution for $f_{2}(Q,B_Q)$ with $\kappa=\lambda=p=1$}\label{fig6}
\end{figure}
\par\end{center}

Hence, considering a minimal Yukawa coupling, the action describing the $1/2$-spin in a $5$-dimensional Dirac field becomes:
\begin{eqnarray}\label{1}
\mathcal{S}_{1/2}=\int \sqrt{-g} \Big(\overline{\Psi}i\Gamma^M D_M\Psi -\xi \phi\overline{\Psi}\Psi\Big)d^5x.
\end{eqnarray}
Here, $\xi$ represents a dimensionless coupling constant. The covariant derivative $D_M$ is expressed as $D_M=\partial_M +\Omega_M$, where $\Omega_M$ denotes the torsion-free spin connection. This connection is defined within the framework of the Levi-Civita connection terms:
\begin{eqnarray}\label{3}
\Omega_M=\frac{1}{4}\Big(\Gamma_M\ ^{{\overline{N}}{\overline{Q}}}\Big)\ \Gamma_{\overline{N}}\Gamma_{\overline{Q}}.
\end{eqnarray}
The curved Dirac matrices $\Gamma^{{M}}$ are derived from the plane Dirac matrices $\Gamma^{\overline{M}}$ and the \textit{vielbeins} $E_{\overline{M}}\ ^M$, following the relation:
\begin{eqnarray}
\Gamma^M=E_{\overline{M}}\ ^M \Gamma^{\overline{M}}.
\end{eqnarray}
These matrices satisfy the Clifford algebra $\{\Gamma^M,\Gamma^N\}=2g^{MN}$. The \textit{vielbeins} establish a tangent space and establish a relationship with the metric through:
\begin{eqnarray}
g_{MN}=\eta_{\overline{M}\overline{N}}E^{\overline{M}}_M E ^{\overline{N}}_N.
\end{eqnarray}
We define the slashed capital Latin indices ($\overline{M},\overline{N},...=0,1,2,3,4$) to represent the coordinates of the tangent space. To simplify, a transformation is applied as $dz=e^{-A(y)}dy$, resulting in the metric taking the form:
\begin{eqnarray}
ds^ 2=e^{2A}(\eta ^{\mu\nu}dx^\mu dx^\nu+dz^2).
\end{eqnarray}

Therefore, the Dirac equation (\ref{1}) can be expressed as:
\begin{eqnarray}\label{7}
\Big[\gamma^{\mu}\partial_\mu+\gamma^4(\partial_z+2\dot{A})-\xi e^A\phi\Big]\psi=0.
\end{eqnarray}
Here, the spinor representation is defined as: 
\begin{eqnarray}
\Psi\equiv\Psi(x,z)=\left(\begin{array}{cccccc}
\psi\\
0\\
\end{array}\right),\ 
\Gamma^{\overline{\mu}}=\left(\begin{array}{cccccc}
0&\gamma^{\overline{\mu}}\\
\gamma^{\overline{\mu}}&0\\
\end{array}\right),\ \Gamma^{\overline{z}}=\left(\begin{array}{cccccc}
0&\gamma^4\\
\gamma^4&0\\
\end{array}\right),
\end{eqnarray}
and (\ $\dot{ }$\ ) denotes a derivative with respect to $z$. 

Through the Kaluza-Klein decomposition of the spinor
\begin{eqnarray}
\psi=\sum_n[\psi_{L,n}(x)\varphi_{L,n}(z)+\psi_{R,n}(x)\varphi_{R,n}(z)],
\end{eqnarray}
we obtain the coupled equations
\begin{eqnarray}\label{9}
\Big[\partial_z+\xi e^A \phi\Big]\varphi_{L}(z)&=&m \varphi_{R}(z),\nonumber\\
\Big[\partial_z-\xi e^A\phi\Big]\varphi_{R}(z)&=&-m \varphi_{L}(z),
\end{eqnarray}
where $\gamma^4\psi_{R,L}=\pm\psi_{R,L}$ represent the left-handed and right-handed components from the Dirac field, and $\gamma^\mu\partial_\mu\psi_{R,L}=m\psi_{L,R}$ holds.

Decoupling of equation (\ref{9}) results in Schr{o}dinger-like equations:
\begin{eqnarray}\label{10}
\Big[-\partial^2_z+V_L(z)\Big]\varphi_{L}(z)&=&m^2 \varphi_{L}(z),\nonumber\\
\Big[-\partial^2_z+V_R(z)\Big]\varphi_{R}(z)&=&m^2 \varphi_{R}(z),
\end{eqnarray}
where $V_{R,L}(z)=U^2 \pm\partial_{z}U$ represents the effective potential, and $U=\xi e^A \phi$ is the superpotential. It is worth noting that this equation takes the form of supersymmetric quantum mechanics (SUSY-type), ensuring the absence of tachyonic Kaluza-Klein (KK) states. Additionally, the supersymmetric structure permits the existence of a well-localized massless mode in the form:
\begin{eqnarray}
\varphi_{R0,L0}(z)\propto e^{\pm\int Udz}.
\end{eqnarray}

Only left-chirality fermions are localized, consistent with what was obtained in GR. For $f_1$, we plot the behavior of the effective potential and the massless fermionic mode in Fig. \ref{fig7}. As we increase the value of $k$ for $n=1$, we observe that the potential tends to become more confining and the massless mode more localized. The same trend holds for $n=2$ and $3$. However, something unexpected occurs: the potential well divides as $k$ increases. This division occurs in the same range of values as the brane split. This indicates that brane splitting interferes with fermion localization. Furthermore, we can conclude that the modification of the gravitational model can make the massless fermion more or less localized in the brane. This assertion is confirmed through the behavior of the effective potential and the massless mode for the $f_2$ model. As we further modify the model, the potential well splits. The fermionic mode senses these changes and becomes more localized but with a flattened peak (Fig. \ref{fig8}).

\begin{center}
\begin{figure}[ht!]
\begin{centering}
\subfloat[$n=1$]{\centering{}\includegraphics[scale=0.6]{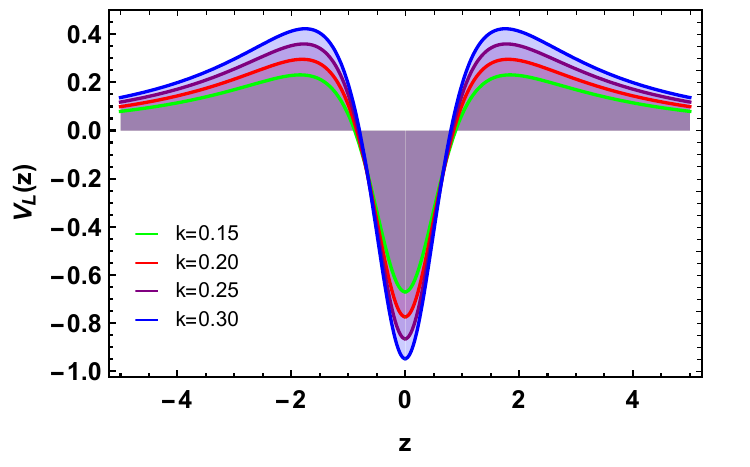}
\includegraphics[scale=0.6]{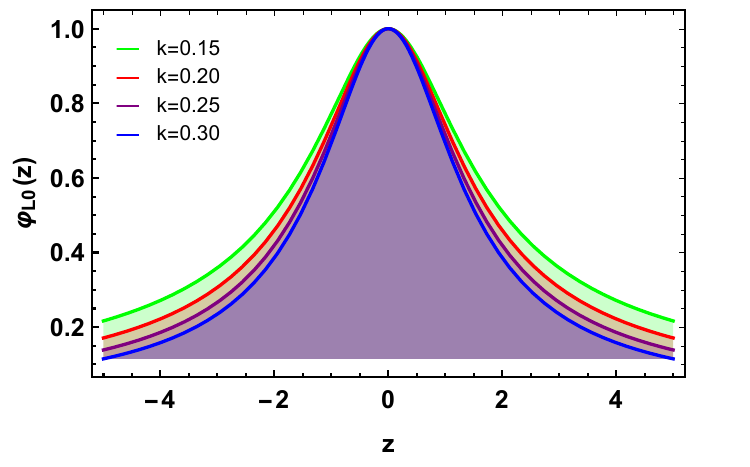}}
\par\end{centering}
\begin{centering}
\subfloat[$n=2$]{\centering{}\includegraphics[scale=0.6]{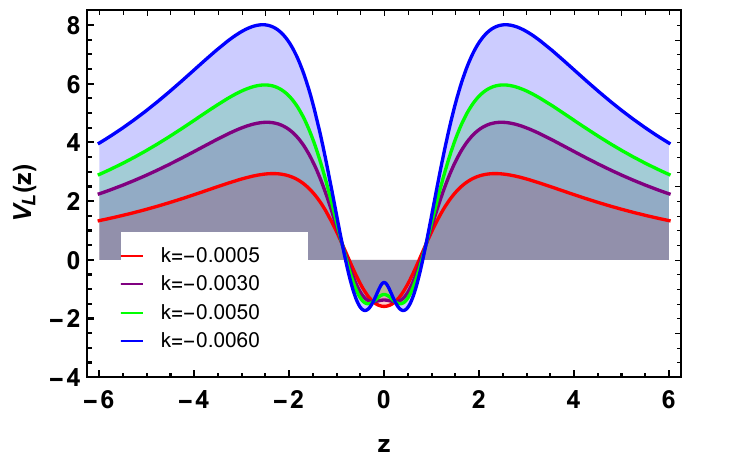}\includegraphics[scale=0.6]{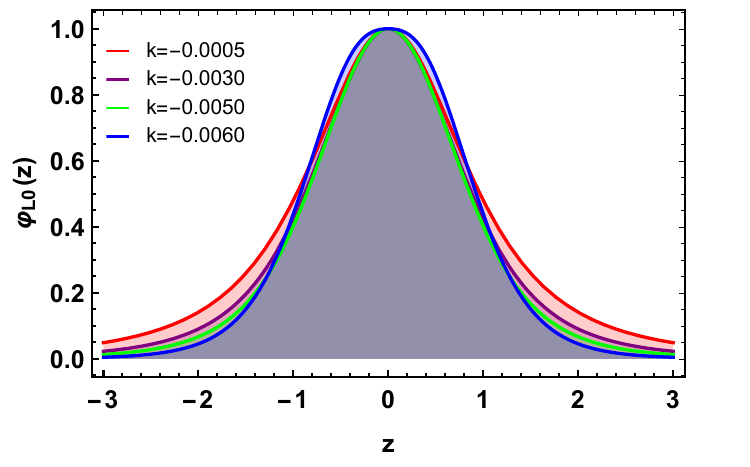}}
\par\end{centering}
\begin{centering}
\subfloat[$n=3$]{\centering{}\includegraphics[scale=0.6]{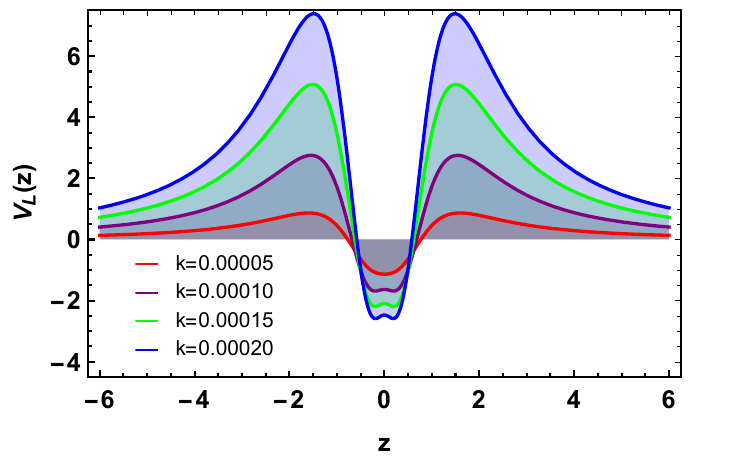}\includegraphics[scale=0.6]{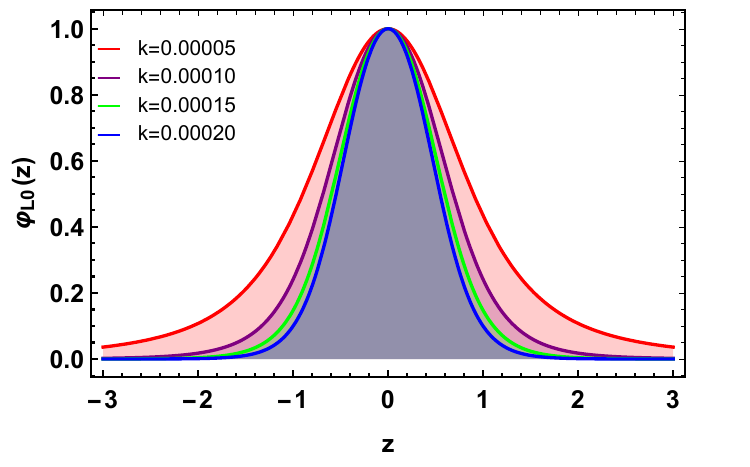}}
\par\end{centering}
\centering{}\caption{Effective potential and solution of massless modes for $f_{1}(Q,B_Q)$ with $\kappa=\xi=\lambda=p=1$}\label{fig7}
\end{figure}
\par\end{center}

\begin{center}
\begin{figure}[ht!]
\begin{centering}
\subfloat[$k_2=-0.001$]{\centering{}\includegraphics[scale=0.6]{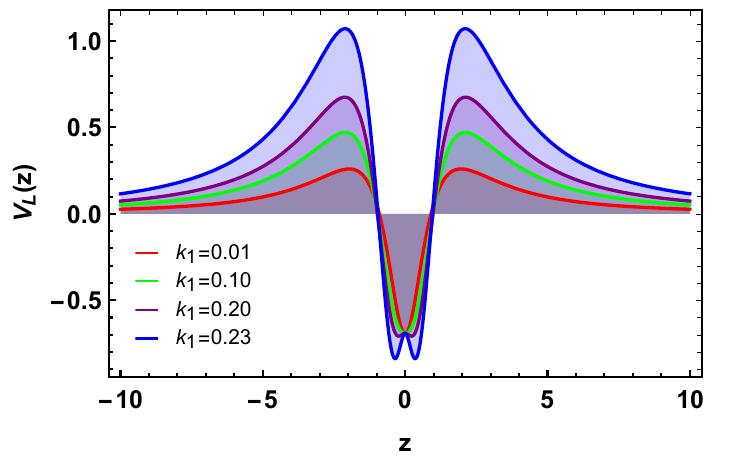}\includegraphics[scale=0.6]{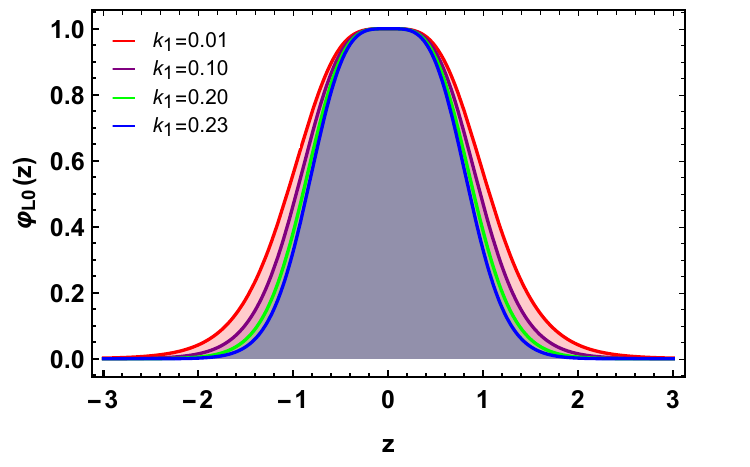}}
\par\end{centering}
\begin{centering}
\subfloat[$k_1=0.01$]{\centering{}\includegraphics[scale=0.6]{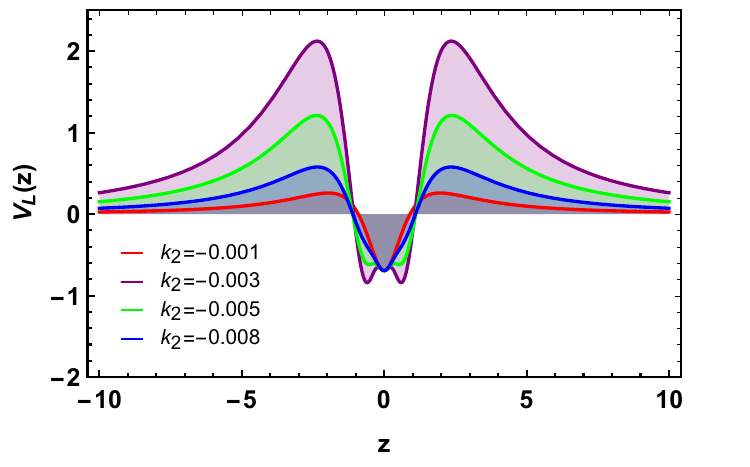}\includegraphics[scale=0.6]{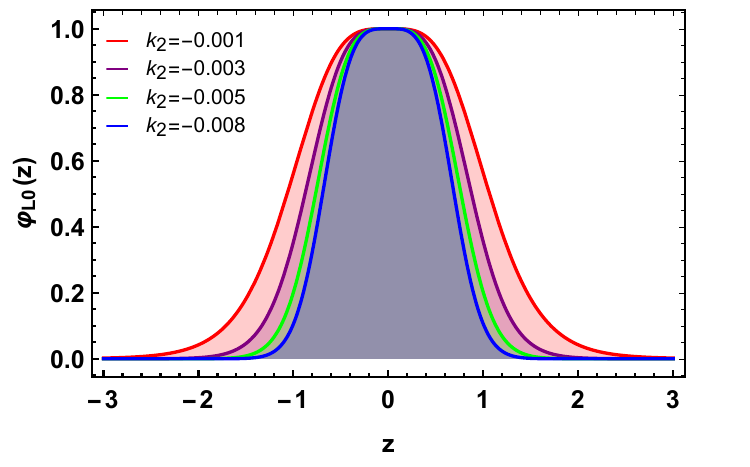}}
\par\end{centering}
\centering{}\caption{Effective potential and solution of massless modes for $f_{2}(Q,B_Q)$ with $\kappa=\xi=\lambda=p=1$}\label{fig8}
\end{figure}
\par\end{center}

Furthermore, we can utilize quantum information measurements to ascertain the conditions that promote fermion localization in our model — that is, the configurations most likely to confine the massless fermion to the brane. Thus, our discussion will delve deeper into the fundamental concepts of Shannon entropy and its application to our specific models \cite{Shannon}.

To define Shannon entropy, we employ the Fourier transform on the massless mode function
\begin{equation}\label{fouu}
\vert\varphi_{L0,R0}(p_z)\vert^{2}=\frac{1}{\sqrt{2\pi}}\int_{-\infty}^{\infty}\vert\varphi_{L0,R0}(z)\vert^{2},\text{e}^{-ipz} dz,
\end{equation}
where $p_z$ signifies the coordinate within momentum space (or reciprocal space). This transformation allows us to delineate Shannon entropy for both position and momentum spaces:
\begin{eqnarray}\label{0.11}
S_{z}&=&-\int_{-\infty}^{\infty}\vert\varphi_{L0,R0}(z)\vert^{2}\ln\vert\varphi_{L0,R0}(z)\vert^{2}dz,\nonumber\\
S_{p_z}&=&-\int_{-\infty}^{\infty}\vert\varphi_{L0,R0}(p_z)\vert^{2}\ln\vert\varphi_{L0,R0}(p_z)\vert^{2}dz.
\end{eqnarray}

These entropy measures yield an uncertainty relation known as the BBM relation \cite{Beckner,Bialy}, named after its proposers Beckner, Bialynicki-Birula, and Mycielski. Notably, this entropic uncertainty relationship serves as a more effective alternative to the Heisenberg uncertainty principle. The BBM uncertainty relation is expressed as:
\begin{equation}
S_{z}+S_{p_z}\geq D(1+\text{ln}\pi).
\end{equation}
Here, $D$ denotes the dimensions capable of perceiving changes in system information. In our model, only the extra dimension possesses the capability to sense the entropic modifications of the system ($D=1$), i.e., $S_{z}+S_{p_z}\geq 2.14473$.

The Shannon information measures are explored numerically through tables \ref{tab1} and \ref{tab2}.

\begin{table}[ht!]
\centering
\caption{
Values of Shannon entropy measurements for $f_{1}(Q,B_Q)$ with $\kappa=\xi=\lambda=p=1$ .\label{tab1}}
\begin{tabular}{|c||c|c|c|c|}
\hline
\hline
$n$ & $k$ & $S_{z}$ & $S_{p_z}$ & $S_{z}+S_{p_z}$ \\ \hline
\hline
1 & 0.15  & 1.06461 & 1.12187  & 2.18648  \\
  & 0.20  & 1.04662 & 1.12789  & 2.17451   \\
  & 0.25  & 1.02956 & 1.13365  & 2.16321 \\ 
  & 0.30  & 1.01336 & 1.13913  & 2.15249  \\ \hline
\hline
2 & -0.0005  & 2.33367 & 0.13141  & 2.46508  \\
  & -0.0030  & 2.29155 & 0.14567  & 2.43722   \\
  & -0.0050  & 2.26880 & 0.18786  & 2.45676 \\ 
  & -0.0060  & 2.21089 & 0.21826  & 2.42915  \\ \hline
 \hline
3 & 0.00005  & 2.12616 & 0.48797  & 2.61413  \\
  & 0.00010  & 1.82405 & 0.57093  & 2.39498   \\
  & 0.00015  & 1.65778 & 0.64169  & 2.29947 \\ 
  & 0.00020  & 1.55157 & 0.74355  & 2.29512  \\ \hline
\end{tabular}\\
\end{table}

\begin{table}[ht!]
\centering
\caption{
Values of Shannon entropy measurements for $f_{2}(Q,B_Q)$ with $\kappa=\xi=\lambda=p=1$ .\label{tab2}}
\begin{tabular}{|c||c|c|c|c|}
\hline
\hline
$k_1$ & $k_2$ & $S_{z}$ & $S_{p_z}$ & $S_{z}+S_{p_z}$ \\ \hline
\hline
0.01  & -0.001  & 2.29165 & 0.79878  & 3.09043  \\
0.10  &         & 2.15574 & 0.92087  & 3.07661   \\
0.20  &         & 1.96040 & 0.97092  & 2.93132 \\ 
0.23  &         & 1.85658 & 1.04169  & 2.89827  \\ \hline
\hline
0.01 & -0.003  & 1.69363 & 1.13691  & 2.83054  \\
     & -0.005  & 1.50861 & 1.28490  & 2.79351   \\
     & -0.008  & 1.26738 & 1.30449  & 2.57188 \\ 
 \hline
\end{tabular}\\
\end{table}

For $f_1$, it's noticeable that as we increase the value of $k$ (the deviation from STEGR), the Shannon measurements in position space decrease, indicating greater certainty about the fermion's location in the brane. Conversely, in momentum space, the Shannon measurement tends to increase, suggesting greater uncertainty in the fermion's momentum. This trend intensifies as we further modify our gravitational model (by increasing the value of $n$). Additionally, the total entropy of the system decreases ($S_{z}+S_{p_z}$) (Tab. \ref{tab1}). The same behavior is observed for $f_2$ (Tab. \ref{tab2}). This leads us to conclude that as our gravitational model deviates from the usual STEGR, the likelihood of locating fermions in the brane increases with minimal loss of information about their location. This finding is significant and clearly indicates that to detect such fermions in our world, we need to consider new phenomenological parameters different from those proposed by GR. Indeed, everything suggests that GR alone is insufficient to explain and describe the world in which we live.

Finally, as the mass of the fermion is linked to its energy, we can infer that the greater its mass, the greater its energy. Consequently, the higher the energy of the fermion, the higher the probability that it will be able to escape from the brane into the bulk. The study of the behavior of massive fermions is of paramount importance for the future detection of these particles at the LHC, thereby proving the existence of extra dimensions. It is noteworthy that some particles with similar behavior have already been detected at the LHC, but as the phenomenon occurred only a few times, it was not sufficient to validate the theory conclusively. All of this underscores that we are progressing in the right direction with our studies.

Localization of massive modes is achievable via numerical methods by imposing boundary conditions \cite{Liu2009,Liu2009a,Moreira20211,Moreira:2021wkj} such as:
\begin{eqnarray}\label{ade}
\varphi_{\text{even}}(0)&=&c, \quad \partial_z\varphi_{\text{even}}(0)=0, \nonumber\\
\varphi_{\text{odd}}(0)&=&0, \quad \partial_z\varphi_{\text{odd}}(0)=c.
\end{eqnarray}
These boundary conditions are selected because of the even nature of the effective potentials $V_{R,L}(z)$. Additionally, conditions (\ref{ade}) ensure that the solutions $\varphi_{R,L}(z)$ display behavior corresponding to either even wave functions $\varphi_{\text{even}}$ or odd wave functions $\varphi_{\text{odd}}$.

We numerically plot the profiles of the massive fermionic modes. Interestingly, these modes exhibit solutions resembling free waves. For $f_1$ (Fig. \ref{fig9}), as we increase the influence of $k$, we observe that the amplitudes of the waves near the brane tend to increase, suggesting possible resonant modes \cite{Moreira:2023pkh}. The same trend is observed for $f_2$ (Fig. \ref{fig10}). This indicates that the massive fermions sense the division of the brane, which tends to confine them but without success. As the fermion's energy is proportional to its mass, higher energy fermions have a greater likelihood of escaping into the bulk. Moreover, besides potentially escaping into the bulk, these fermions may also interfere with gravitational wave measurements obtained by the Laser Interferometer Gravitational-Wave Observatory (LIGO), as they exhibit modes resembling free waves.

\begin{center}
\begin{figure}[ht!]
\begin{centering}
\subfloat[$n=2$]{\centering{}\includegraphics[scale=0.6]{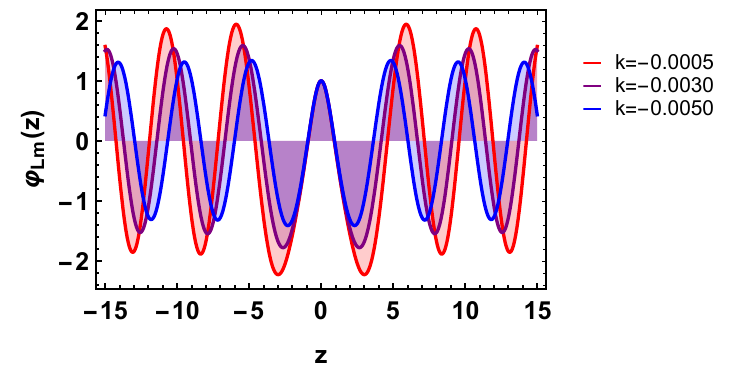}\includegraphics[scale=0.6]{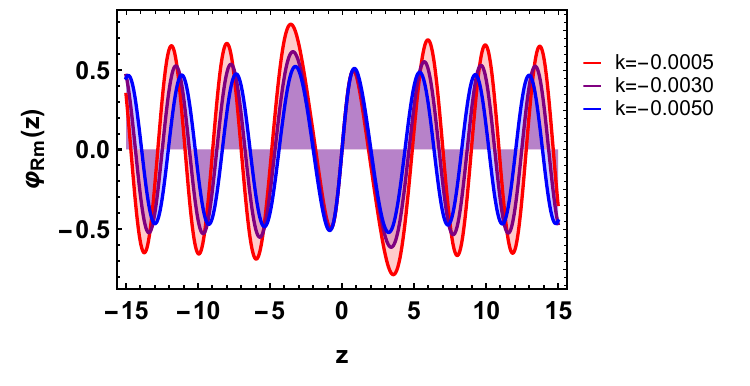}}
\par\end{centering}
\begin{centering}
\subfloat[$n=3$]{\centering{}\includegraphics[scale=0.6]{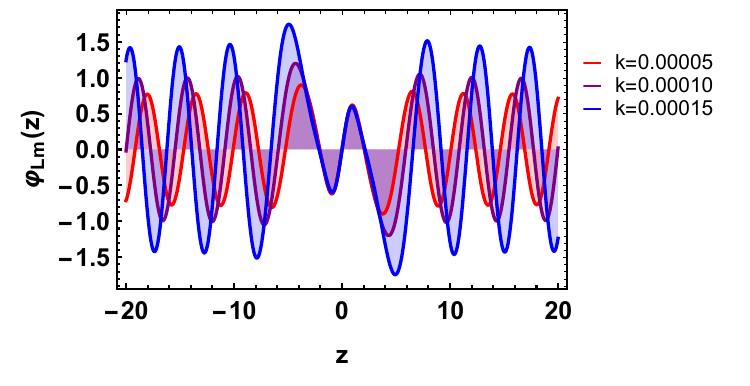}\includegraphics[scale=0.6]{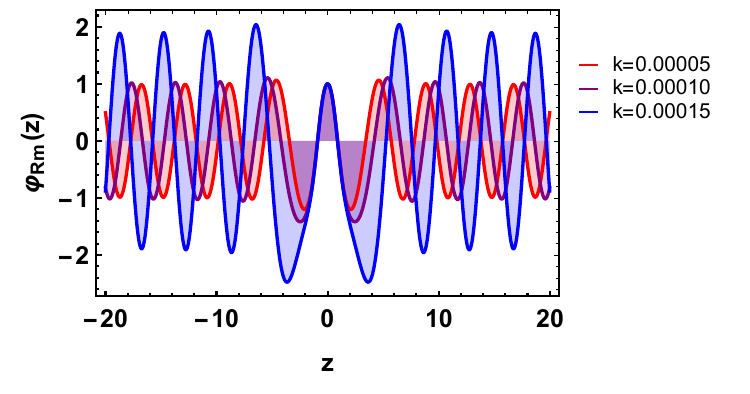}}
\par\end{centering}
\centering{}\caption{Massive mode solutions for $f_{1}(Q,B_Q)$ with $\kappa=\xi=\lambda=p=1$}\label{fig9}
\end{figure}
\par\end{center}

\begin{center}
\begin{figure}[ht!]
\begin{centering}
\subfloat[$k_2=-0.001$]{\centering{}\includegraphics[scale=0.6]{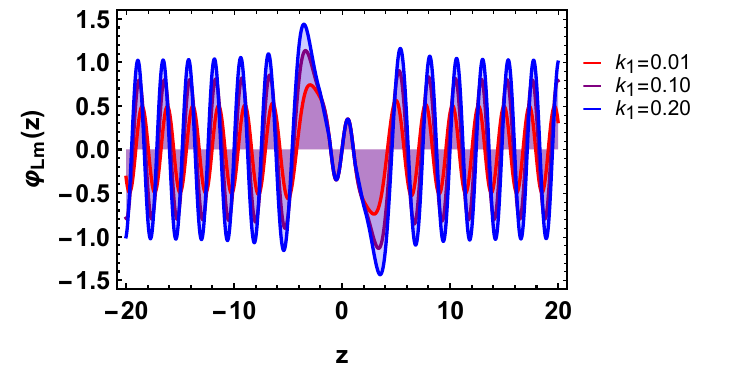}\includegraphics[scale=0.6]{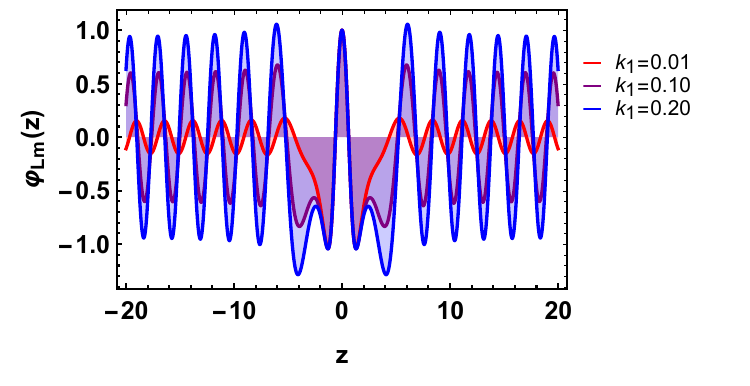}}
\par\end{centering}
\begin{centering}
\subfloat[$k_1=0.01$]{\centering{}\includegraphics[scale=0.6]{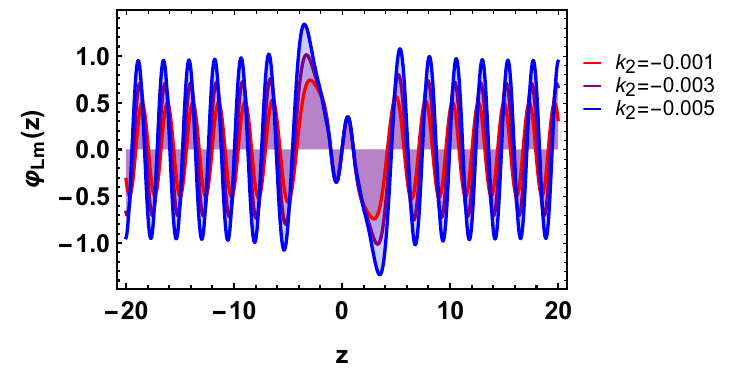}\includegraphics[scale=0.6]{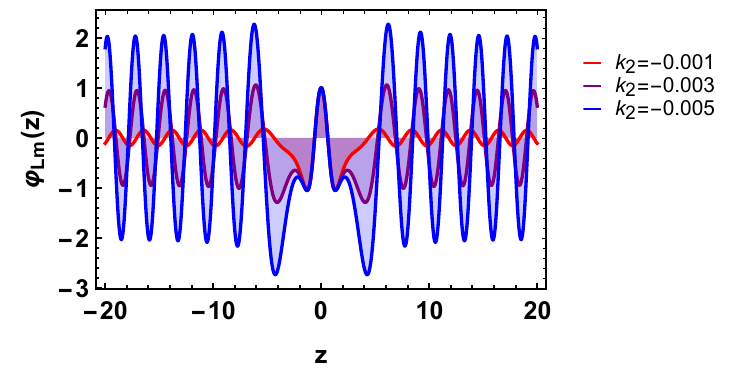}}
\par\end{centering}
\centering{}\caption{Massive mode solutions for $f_{2}(Q,B_Q)$ with $\kappa=\xi=\lambda=p=1$}\label{fig10}
\end{figure}
\par\end{center}

\section{Final remarks}
\label{s3}

In conclusion, armed with our extensive findings, we embark on a comprehensive exploration of braneworld behavior within modified gravity models. To guide our inquiry, we pose fundamental questions and endeavor to address them systematically. First and foremost, we scrutinize the influence of gravitational alterations on brane structure and the behavior of the underlying scalar field. Secondly, we seek to ascertain the most plausible gravitational configurations that may characterize our universe. Thirdly, we probe the impact of gravitational modifications on the spatial distribution of particles, such as spin-$1/2$ fermions, crucial for experimental validation of extra dimensions.

Our analysis reveals intriguing insights, depicted graphically for clarity. In the case of the first gravitational model, distinct energy density peaks emerge with deviations from the standard scenario, indicative of a split in the brane. Moreover, utilizing the DCE, we discern stable configurations coinciding with the brane split intervals, affirming their likelihood in our models. Similarly, for the second gravitational model, analogous observations reinforce the association between stable configurations and brane division. Furthermore, our examination extends to fermion localization within the brane, necessitating careful consideration of scalar field characteristics. Notably, left-chirality fermions exhibit localization, influenced by the gravitational model's deviations from the usual STEGR. Numerical simulations elucidate the emergence of domain walls and phase transitions within the brane, indicative of gravitational model modifications.

Delving deeper, we leverage quantum information measurements, particularly Shannon entropy, to assess fermion localization probabilities. Strikingly, as the gravitational model diverges from the standard, the certainty of fermion localization within the brane increases, suggesting the inadequacy of GR to fully explain our universe. Lastly, we explore the implications of massive fermions, crucial for future experimental endeavors, such as those conducted at the LHC. Our numerical simulations elucidate wave-like modes of massive fermions, highlighting their potential to escape the brane and interfere with gravitational wave measurements.

In essence, our comprehensive investigation underscores the necessity of considering modified gravitational models to accurately describe the complex dynamics of our universe. It not only challenges the conventional framework of GR but also offers promising avenues for future theoretical and experimental explorations, steering us closer to a comprehensive understanding of our cosmos.

\section*{Acknowledgments}S.H. Dong acknowledges the partial support of project 20240220-SIP-IPN, Mexico, and commenced this work on a research stay in China. M.E.R. expresses gratitude to Conselho Nacional de Desenvolvimento Científico e Tecnológico - CNPq, Brazil, for partial financial support.

\end{document}